\documentclass[sigconf]{acmart}
\usepackage{graphicx,epsfig,amsmath,amssymb,bm}
\usepackage{amssymb,balance}
\usepackage{amsmath}
\usepackage{amsthm}
\usepackage{tikz}
\usepackage{amssymb}

\newfont{\bbb}{msbm10 scaled 500}

\usepackage{bbm}
\usepackage{tabularx}
\usepackage{multirow}
\usepackage{hhline}
\usepackage{verbatim}

\usepackage{subcaption}

\usepackage{color}   

\newcolumntype{M}[1]{>{\centering\arraybackslash}m{#1}}
\newcommand{\beqa}{\begin{eqnarray}}
\newcommand{\eeqa}{\end{eqnarray}}
\newcommand{\dsp}{\displaystyle}

\usepackage[ruled,linesnumbered]{algorithm2e}

\setcopyright{rightsretained}

\begin{document}

\title{Signal Jamming Attacks Against Communication-Based Train Control: Attack Impact and Countermeasure}
\titlenote{This work was supported by the National Research
Foundation (NRF), Prime Minister's Office, Singapore, under
its National Cybersecurity R\&D Programme (Award
No. NRF2014NCR-NCR001-31) and administered by the
National Cybersecurity R\&D Directorate.}

\copyrightyear{2018} 
\acmYear{2018} 
\setcopyright{acmcopyright}
\acmConference[WiSec '18]{Proceedings of the 11th ACM Conference on Security \& Privacy in Wireless and Mobile Networks}{June 18--20, 2018}{Stockholm, Sweden}
\acmPrice{15.00}
\acmDOI{10.1145/3212480.3212500}
\acmISBN{978-1-4503-5731-9/18/06}

\author{Subhash Lakshminarayana}
\affiliation{%
  \institution{Advanced Digital Sciences Center, \\ Illinois at Singapore, Singapore} 
}
\email{subhash.l@adsc-create.edu.sg}

\author{Jabir Shabbir Karachiwala}
\affiliation{%
  \institution{Advanced Digital Sciences Center, \\ Illinois at Singapore, Singapore}
}
\email{jabir.k@adsc-create.edu.sg}

\author{Sang-Yoon Chang}
\affiliation{
  \institution{University of Colorado Colorado Springs}
 }
\email{schang2@uccs.edu}

\author{Girish Revadigar}
\affiliation{
  \institution{Singapore University of Technology and Design, Singapore}
}
\email{girish\_shivalingappa@sutd.edu.sg}

\author{Sristi Lakshmi Sravana Kumar}
\affiliation{
  \institution{Advanced Digital Sciences Center, \\ Illinois at Singapore, Singapore}
}
\email{sravana.s@adsc-create.edu.sg}

\author{David K.Y. Yau}
\affiliation{%
  \institution{Singapore University of Technology and Design, Singapore}
  }
\email{david\_yau@sutd.edu.sg}

\author{Yih-Chun Hu}
\affiliation{
  \institution{University of Illinois at Urbana-Champaign}
}
\email{yihchun@illinois.edu}

\renewcommand{\shortauthors}{S. Lakshminarayana et.al.}
\renewcommand{\shorttitle}{Signal Jamming Attacks Against CBTC: Attack Impact \& Countermeasure}

\addtolength{\textfloatsep}{-5.5mm}
\addtolength{\floatsep}{-4mm}

\begin{abstract}
We study the impact of signal jamming attacks against the communication based
train control (CBTC) systems and develop the countermeasures to limit the attacks' impact.
CBTC supports the train operation automation and moving-block signaling, which improves the transport efficiency. We consider an attacker jamming the wireless communication between the trains or the train to wayside access point, which can disable CBTC and the corresponding benefits.
In contrast to prior work studying jamming only at the physical or link layer, we study the real impact of such attacks on end users, namely train journey time and passenger congestion. Our analysis employs a detailed model of leaky medium-based communication system (leaky waveguide or leaky feeder/coaxial cable) popularly used in CBTC systems. To counteract the jamming attacks, we develop a mitigation approach based on frequency hopping spread spectrum taking into account domain-specific structure of the leaky-medium CBTC systems. Specifically, compared with existing implementations of FHSS, we apply FHSS not only between the transmitter-receiver pair but also at the track-side repeaters. To demonstrate the feasibility of implementing this technology in CBTC systems, we develop a FHSS repeater prototype using software-defined radios on both leaky-medium and open-air (free-wave) channels. We perform extensive simulations driven by realistic running profiles of trains and real-world passenger data to provide insights into the jamming attack's impact and the effectiveness of the proposed countermeasure.
\end{abstract}

%


%



 
\maketitle

\section{Introduction}
With rapid explosion in urban populations, metro systems around the world are getting increasingly
congested. Information and communication technologies (ICTs) can play a key role in relieving the congestion by improving the railway infrastructure utilization, and are being increasingly adopted by railway operators. However, their adoption also makes railways vulnerable to cyber attacks. 
Existing cybersecurity of modern railways typically appeals to air gaps that isolate the ICT systems from public networks. However, there are growing instances of
successful air-gap breaches in railways \cite{UK,Boston} and other critical infrastructures 
(e.g., Black Energy and the Stuxnet attacks \cite{karnouskos2011}, \cite{Ukraine2016}). 
Such security breaches can have severe consequences on end users. This is particularly true of railways, due to deep involvement of humans who use them everyday in large numbers, where physical isolation is highly questionable in the first place.

In this paper, we study the cybersecurity of communication-based train control (CBTC) \cite{FarooqCBTC2017}, an automatic train control system that enables trains to run with shorter headways, thereby improving track utilization. In CBTC, the trains can continuously exchange their states of motion (i.e., location, velocity, and acceleration/deceleration capabilities) among each other over high-speed wireless communication links, and optimize their headway accordingly. However, CBTC has stringent requirements for communication availability, and the loss of communication can lead to severe disruptions. A recent real-world incident occurred for the Singapore metro \cite{CircleLine}, in which a train with faulty signaling hardware affected the communication of other trains traveling in its vicinity. This resulted in the trains activating their emergency brakes unnecessarily, leading to multiple delays and widespread disruptions. Another incident involving CBTC signaling fault resulted in more serious train collision \cite{EWCollision}. These incidents highlight the importance of understanding cyber attacks that can cause the loss of signaling in CBTC and developing countermeasures.

We consider signal jamming attacks against the CBTC, in which the attacker injects an interference signal into the wireless transmission in order to disrupt the communications (specifically, train-to-train or train-to-trackside-infrastructure communications). The jamming can disable the CBTC and negate its benefits such as transport efficiency. The threat is acute in urban train systems as they are accessed by and share the same physical space with the public, as opposed to other critical infrastructures that might be physically isolated.
This co-location heightens the risk as say rogue attackers close to their targets may readily impart strong interferences. They can readily do so as outsiders; there is no need for prior compromise of any credential systems. Moreover,  the availability of software-defined radio (SDR) has lowered the bar significantly for would-be attackers. 
They can now launch jamming attacks by simply commanding a software API, without much expertise in low-level radios and signal processing. As a result, jamming has emerged as a major focal point of cybersecurity concerns for train systems~\cite{deniau2014,apta2014,FarooqCBTC2017}.

The attacker's ability to jam the train communications critically depends on  
propagation characteristics of the wireless medium in question. In this work, we focus on the paradigm of \emph{leaky-medium} communication (using waveguide and coaxial cable), which is popular for trains due to their constrained mobility by the railway tracks \cite{Leaky2009}. 
To support communications over long distances, the leaky medium-based communication typically employs a tandem of \emph{repeaters} to compensate for path loss. We address the impacts of the leaky medium as an understudied subject compared with traditional \emph{free-wave} channels.

We aim to answer the following two research questions in this paper.  
\emph{(1) How to quantify the true impact of signal jamming attacks?} 
Analysis of the impact will underline the development of countermeasures and evaluation of their effectiveness. 
However, the analysis is challenging because railways are complex cyber-physical systems. 
They involve a number of interdependent subsystems operating in concert. In our setting, for example, the trains' wireless communications impact their motion (e.g., velocities and corresponding headways), which in turn affects passenger flows (i.e., passenger wait times and congestion). The latter metrics are critical since they measure the effects on end users and stakeholders who truly matter in everyday applications. 
The investigation must capture these interdependencies and expose the truly important end performance of the CBTC.

\emph{(2) How to design effective countermeasure against the jamming attack in today's train systems?}
The CBTC environment provides not only unique challenges but also opportunities for security. We take advantage of the CBTC communication architecture to achieve increased jamming resistance compared with prior work.

In addressing the above two research questions, we make the following main contributions. \\
We analyze jamming in leaky medium-based communication, which extends prior such results 
for the free-wave medium (see also the discussions in Sec.~\ref{sec:Related}) and is critical for the CBTC application domain. The leaky medium has channel characteristics quite distinct from the case of free wave. 
Whereas recent work \cite{Sang-Yoon2015} has pointed out specific issues of leaky medium-based communications under jamming attacks,  
their evaluations are limited to the physical communication layer only. 
Importantly, we evaluate the impacts of the affected train operations from the end user's perspective, namely wait times and congestion. 
Clearly, train operators are primarily concerned about the service they provide ultimately to their customers, instead of any low level details of the communications per se. 
It is because the service quality affects directly their reputation and profitability.  
Moreover, operators usually face significant financial penalties imposed by governments for service disruptions or delays. Their licence may even be revoked in extreme cases. For instance, the U.K. has a penalty scheme that holds rail operators directly responsible for significant service problems \cite{RailLate}. The challenge for understanding the end impacts of CBTC jamming attacks is the lack of an evaluation platform that can integrate the diverse interacting components operating at different layers of the overall system. To meet the challenge, in this work we developed a co-simulation platform that admits holistically (i) a model of train motion under different signaling modes, (ii) a model of the leaky medium-based wireless communications that affect the signaling and train control in turn, and (iii) incorporation of real-world passenger flow datasets \cite{Zhang:2015} that specify the levels and patterns of demand that are key to the system's performance in real operation.

The results reveal that while jamming in free-medium communication may have impact over a limited range only, due to natural signal attenuation, jamming in leaky-medium communication can be impactful throughout the train communication space. This is because jamming over leaky medium can leverage the signal amplifications of the repeaters to extend its effects over much longer distances. For instance, our co-simulations show that the presence of a single jammer can increase the train journey time by up to $40~$minutes, which is approximately a $35\%$ increase, and the average passenger journey time (sum of waiting time and the travel time) by about $15$~minutes. Comparatively, jamming over the free-wave medium will only have negligible impact, i.e., less than $1~$ minute increase of train journey time. These results highlight the risk of jamming attacks on train operation in realistic deployments.

Second, we propose and evaluate a defense measure to mitigate the jamming attacks. 
Our defense builds on frequency hopping spread spectrum (FHSS)~\cite{pickholtz1982,simon1994,simplemac_2012,strasser2008}
to protect the availability of the CBTC communications. In FHSS, the legitimate parties randomize the frequency channel access for transmission, so that the selected channels are dynamic and will appear random to the attackers. 
In this work, we improve the effectiveness of the FHSS by exploiting the domain-specific repeater-based communication structure of the CBTC. Specifically, in contrast to prior work, 
we employ the FHSS not only at the source-destination pair but also the repeaters between them.

We demonstrate the feasibility of implementing the proposed defense in CBTC using an SDR-based leaky-coaxial cable testbed. To realize the anti-jamming at the strategic CBTC repeaters, we develop a novel \emph{FHSS repeater} prototype. Based on the setup of real-world train systems, we consider a trusted entity (e.g., the train control center) that can communicate securely with the transmitter, receiver, and repeaters to control the overall operation. We conduct extensive experiments on the testbed to evaluate prototype. The results show that the signal-to-interference-noise ratio (SINR) at the receiver can be significantly enhanced by adopting the FHSS mitigation. Moreover, using our co-simulator platform, we show that the impact of jamming attacks on the train journey time and passenger congestion can be substantially reduced by implementing the FHSS with $10$ channels.

The rest of the paper is organized as follows. Section~\ref{sec:Related} reviews
 related work. Section~\ref{sec:CBTC_Overview} presents an overview of the CBTC communication system and the jamming threat. Section~\ref{sec:Train_Motion} models the train motion. 
The wireless communication model and the attacker model are presented in Section~\ref{sec:Wireless}.
The proposed FHSS-based mitigation is presented in Section~\ref{sec:Mitigation}. Section~\ref{sec:Sim_Res} reports our
simulation results and Section~\ref{sec:FHSS_proto} presents the FHSS repeater prototype. Section~\ref{sec:Conc} concludes.

\section{Related Work}
\label{sec:Related}
With widespread adoption of CBTC in urban metros, recent research 
has optimized the train motion profile leveraging on accurate tracking of the next train or obstacle, e.g., \cite{Wang_ECC2013, YaoPTC2015, ZhuCBTC2014, WangCognitive2015, ZhaoRob2015}. 
However, none of the aforementioned work on CBTC has addressed it from a cybersecurity perspective. The security problem is imperative, since modern metros integrate ICT increasingly and are critical infrastructures that attract
attacks. Emerging work on railway cybersecurity has focused on trains' traction power control \cite{LakshDSN2016}, whereas we focus on a different kind of attacks (i.e., jamming) against a very different functional module (i.e., CBTC).

Jamming is a widely recognized concern for wireless systems and has been well studied
\cite{simplemac_2012,basar1983,strasser2008,firouzbakht2012,vohuu2016}. Solutions have been proposed that use spread spectrum technology to increase interference resistance, e.g.,~\cite{pickholtz1982,simon1994,simplemac_2012,strasser2008}. 
Existing work has focused predominantly on jamming in free-wave medium only, however. To complement the state of the art, we provide an in-depth study of jamming in leaky medium-based communications for the context of train communications particularly. 
Our work is related to a recent result showing that jamming can gain power through leaky waveguides and repeaters~\cite{Sang-Yoon2015}. But we make contributions beyond the prior work in two important respects. First, we develop a non-trivial co-simulator to evaluate the true end impacts of the attack, whereas they report results for the physical communication layer only. Second, they do not provide defense measures against the attack, whereas we design an FHSS-based defense and prototype it on an SDR platform. Using our co-simulator, we likewise provide novel results on the effectiveness of the defense from an end user's perspective.

\section{Overview of CBTC and Jamming Threat}
\label{sec:CBTC_Overview}
Many modern-day metro systems support Automatic Train Operation (ATO) to enable the automation of train operations. The train communication consists of two phases: a wireless part between the vehicles and the wayside access point and an internal wired network between the wayside access point and the centralized Operational Control Center (OCC). To realize greater efficiency, CBTC leverages on train-to-train and train-to-wayside access point communication. The train localization signaling is based on the interaction of the vehicles with the beacons or transponders (placed along the railway track), called \emph{balise}, that detects the presence of the vehicles. Thereafter, the wayside access point relays the vehicle location to the remote OCC systems via wired connection (e.g., optical cable) in real-time, so that the OCC can monitor the vehicle operation. To take active measures, OCC relays the operational commands to wayside access point via the internal network connection, which in turn wirelessly relays the mission-critical messages to the vehicles. Examples of wireless protocols used are Global System for Mobile Communications: Railway (GMS-R) \cite{GSMR} and Terrestrial Trunked Radio (TETRA) \cite{Tetra}. 

\begin{figure}[!t]
\centering
\includegraphics[width=0.48\textwidth,trim={0 1cm 0 1cm}]{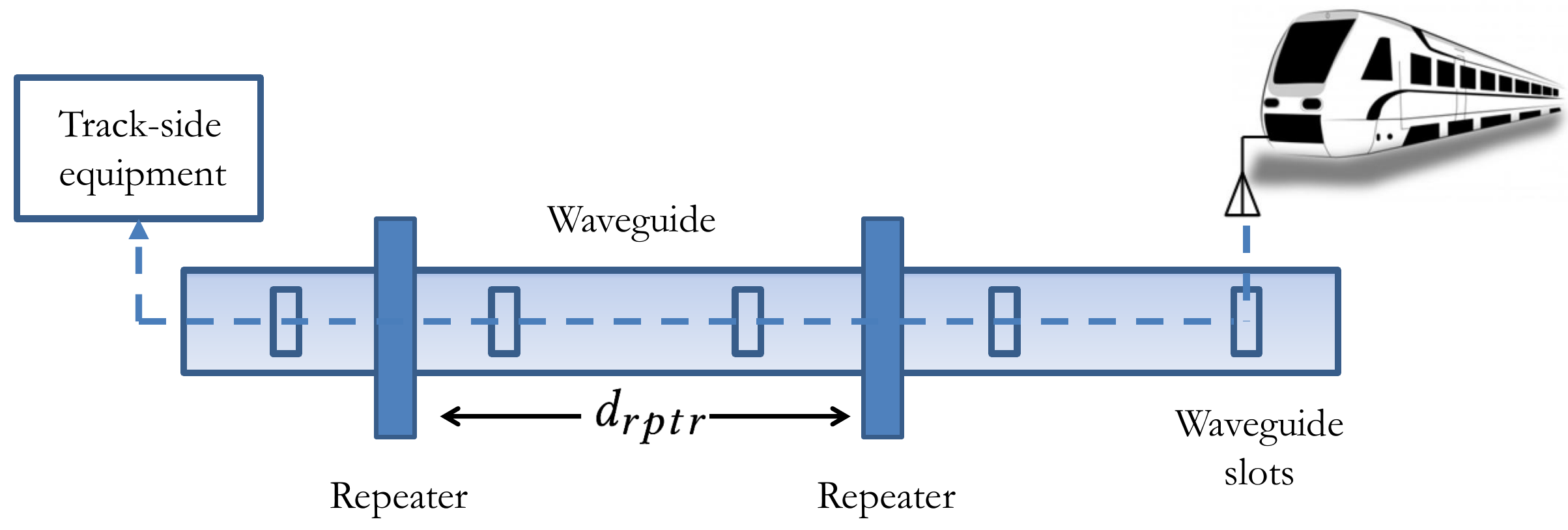}
\caption{Leaky-medium-based train communication.}
\label{fig:Leaky_Med}
\end{figure}

\subsection{Leaky Communication Infrastructure for Train CBTC}
The leaky communication infrastructure, illustrated in Fig.~\ref{fig:Leaky_Med},  
is a communication medium that guides the wireless communication signal. Compared to free-medium communication, the leaky medium limits the signal propagation attenuation and supports a longer signal transmission range. 
These leaky communication structures are in the form of concrete metals with hollow interior (leaky waveguide) or coaxial cables (leaky feeder/coaxial cable), which have well-placed slots to support signal propagation 
to outside of the structures. The wireless/mobile clients (the mobile train and the trainborne transmitter/receiver in our case) 
are not in physical contact with the medium itself, but 
communicate via the signals that are \emph{leaked} from these slots.
The leaked signals attenuate relatively quickly compared to the signal inside the medium 
because there is no longer a physical medium to guide their propagation (open-air medium).


Train communications are ideal to use such mediums 
because the train's mobility is pre-defined and limited to the railway tracks 
(the leaky communication infrastructure is close to the railway tracks, 
and the limited scope of the signal propagation laid-out 
by the leaky infrastructure is appropriate
because the train is never too far away from the railways)
and because the train travels a long distance
(and therefore the signal propagation also needs to support 
a longer distance than what is allowed by a typical omnidirectional signal transmission 
in an open-air medium where the signal propagates in all directions). 
To further support greater communication scope in distance, 
the train communication also implements repeaters 
along the railway-parallel communication infrastructure. 
These repeaters amplify the signal from one side 
and re-transmit it to the other side of the leaky medium.
Therefore, train communications widely adopt such leaky mediums with repeaters
and their performances and benefits for train communications are extensively studied, e.g.,
~\cite{heddebaut2009,heddebaut1990,kawakami1959,wang2013}. 
Section~\ref{sec:Wireless} models the leaky medium for train communications; 
our model uses variables to abstract away from 
the system- and implementation-specific details
and is therefore generally applicable.

\subsection{Threat Model}
We consider an adversary disabling the mission-critical train communications by jamming the train-to-train and train-to-wayside access point wireless communication links. Jamming is typically a low-barrier threat to execute since it does not require a-priori compromise of the communication medium due to the inherent open-nature of the wireless medium. 

We assume that the jammer is within the signal/transmission range of the train communications. For instance, the jammer either has a high-gain antenna or is in relative proximity to the train communication (e.g., along the train tracks or on-board the train). This threat is even more relevant for leaky-medium communication since the jammer can make an impact 
as long as it can get close enough to \emph{any} point along the leaky medium (i.e., the waveguide slots or the repeaters)~\cite{Sang-Yoon2015}, as opposed to the free-medium communication channel (in which the jammer must be in the signal reception range of the receiving entity).

We assume a worst-case scenario with a limited-power, unlimited-energy attacker; thus the jammer continuously transmits the jamming signal (i.e., constant jamming). However, the framework presented in this work can can also admit alternate jamming strategies (e.g., random/reactive jamming). 
Jamming results in SINR degradation and prevents the receiver from retrieving the legitimate signal. Thus, any digital security measures after the received signal demodulation/decoding (e.g., those based on cryptography or network security)  become irrelevant and ineffective. For this reason, existing CBTC protocols place stringent requirements on the legitimate signal's power at the receive antenna as it directly relates to the bit/packet error rate of the train communication \cite{FarooqCBTC2017}.

In the following section, we formally quantify this attack impact by modeling train motion under normal mode of operation and during a communication disruption event.  

\section{Modeling the Train Motion}
\label{sec:Train_Motion}
\begin{figure}[!t]
\centering
\includegraphics[width=0.4\textwidth,trim={0 1cm 0 0}]{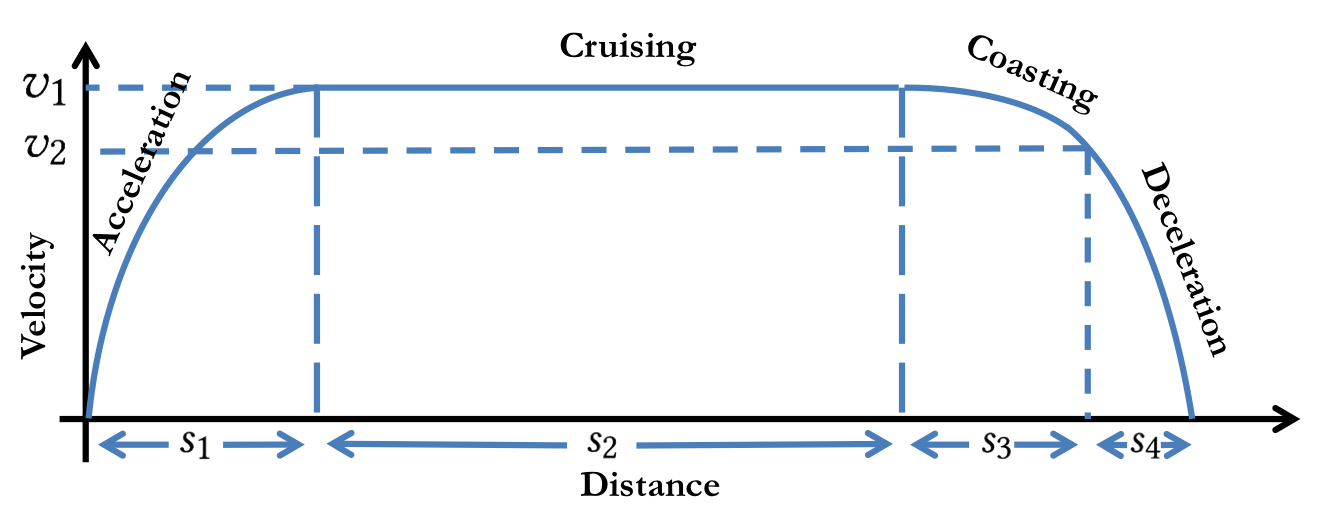}
\caption{Train velocity profile.}
\label{fig:VelocityProfile}
\end{figure}

We begin by modeling the train motion. We distinguish between the planning and the operational phases. During the planning phase, the trains compute a guidance trajectory, i.e., velocity and the acceleration/deceleration profile, which they intend to follow during their journey (see Fig.~\ref{fig:VelocityProfile}). 
During the operational phase, the trains follow the computed guidance trajectory.
However, when an abnormal event occurs
(e.g., if the leading train is too close or if there is a loss of the train communication for a prolonged duration of time), they take a short-term corrective action (specified in Section~4.2). Subsequently, the trains recompute their guidance trajectory for the remaining distance of their journey. The details are presented in the following.

\subsection{Planning Phase}
As in prior work \cite{Wang_ECC2013, YaoPTC2015, ZhuCBTC2014, WangCognitive2015, ZhaoRob2015}, we divide the train's guidance trajectory into four phases -- acceleration, cruising, coasting and braking as shown in Fig.~\ref{fig:VelocityProfile}. 
We denote the train's acceleration and deceleration (due to service brake) by $\alpha$ and $\beta_{\text{ser}}$ respectively. During the cruising phase, the train maintains a constant velocity. 
During coasting, the train applies no traction or braking force, and thus, decelerates due to friction only, which we denote by $a_{fr}$. The duration of acceleration, cruising, coasting and braking phases, which we denote by $T_{1},T_{2},T_{3}$ and $T_{4}$ respectively, are computed to 
minimize the total journey time. We omit the details here and present them in the Appendix~A. 
The train's guidance acceleration profile, which we denote by $a_{\text{plan}}(\tau),$ is then given by
\eqref{eqn:plan_profile} (see Appendix~A), where $t$ is the time index when the guidance trajectory is computed ($t = 0$ at the beginning to train's journey). We denote \mbox{$\mathcal{A}_{\text{plan}} (t) =  \{a_{\text{plan}}(\tau) \}^{t+ T_{1} + T_{2}+T_{3} + T_{4}}_{\tau = t}.$} 
In extreme cases (see Sec.~4.3), the train can stop by applying the emergency brake whose deceleration is denoted by $\beta_{\text{emerg}},$
where $\beta_{\text{emerg}} > \beta_{\text{ser}}.$

\subsection{Operational Phase}
Next, we describe the train motion during its operational phase. We let $v_{l} (t),v_f(t),$ $a_l(t),a_f(t),$ and $s_l (t),s_f(t)$ denote the velocity, acceleration and the position (with respect to the origin) of the leading and following trains respectively at time $t$. For simplicity, we assume that the trains make operational decisions (i.e., whether to accelerate, cruise, coast or decelerate) at discrete time intervals indexed by $t = 0, \Delta t, 2 \Delta t,  3 \Delta t,\dots,$ where $\Delta t$ is the time interval between the decisions. 
We assume that the train's velocity and acceleration remain constant between $n \Delta t$ and $(n+1) \Delta t.$ Under these assumptions, the following/leading train's velocity and position at time $t = (n+1) \Delta t$ can be recursively computed as
\begin{align}
v_i((n+1) \Delta t) &=  v_{i}(n \Delta t)  +a_i(n \Delta t) {\Delta{t}}, \label{eqn:vel_update} \\
s_i((n+1)\Delta t) & = s_i(n\Delta t)+ v_{i} (n\Delta t) \Delta t  +\frac{1}{2} a_i(n\Delta t)  (\Delta{t})^2, \label{eqn:pos_update}
\end{align}
where $v_i(0) = 0, s_i(0) = 0$ and $i = \{f,l\}$. The aforementioned train's operation decision depends on the mode of signaling mode, which we describe in the following.

\begin{figure}[!t]
\centering
\begin{subfigure}{0.23\textwidth}
\includegraphics[width=0.98\textwidth]{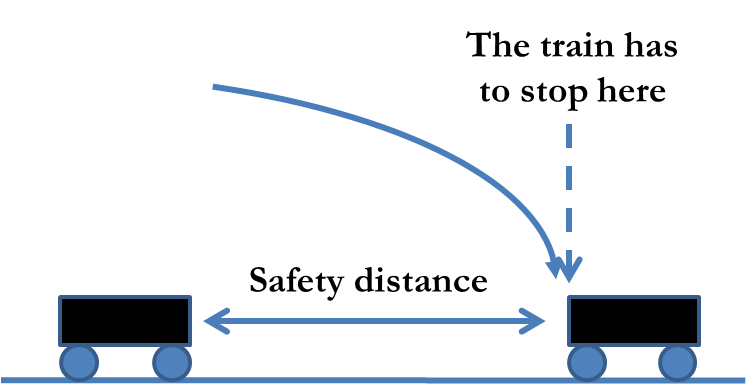}
\end{subfigure}
\begin{subfigure}{0.23\textwidth}
\vspace{0.5cm}
\includegraphics[width=0.98\textwidth]{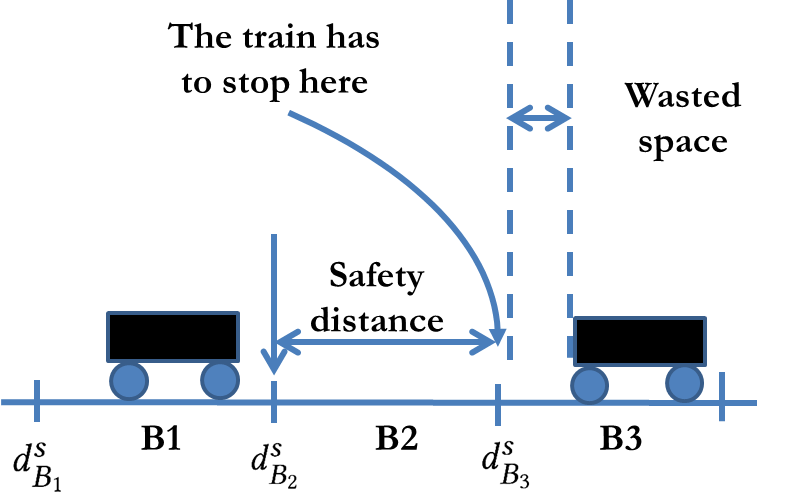}
\end{subfigure}
\caption{Train motion under moving-block (left) and fixed-block (right) signaling modes.}
\label{fig:movingblk}
\end{figure}

{\bf Moving Block Signaling:} Under CBTC, during normal operating conditions, the trains follow the \emph{moving-block signaling} (MBS) mode (refer to Fig.~\ref{fig:movingblk}~(left)), in which the following train
computes a \emph{dynamic headway} based on the state of motion of the leading train and adjusts its velocity accordingly. We let $\widehat{v}_l(t),\widehat{a}_l(t),\widehat{s}_l(t)$ denote the velocity, acceleration and position respectively of the leading train that is communicated to the following train at time $t$  (the detailed description of the communication model is deferred to Section~\ref{sec:Wireless}). Based on this information, the following train computes 
a dynamic headway, denoted by $H(t)$, which is the minimum separation between the two trains to avoid collision under the worst-case stopping scenario (i.e., when the leading train stops by applying the emergency brake). The details of the dynamic headway computation are omitted here and presented in Appendix~B.

{\bf Fall-Back Signaling:} 
To accommodate for the loss of communication, operators have recently started including a \emph{fall-back signaling} mode, in which the trains automatically switch to a fail-safe mode of operation, which ensures that the trains do not collide \cite{Fallback2}, \cite{DLTSignal}.  In this work, we assume that the fall-back mode corresponds to the \emph{fixed-block signaling} (FBS) system (Fig.~\ref{fig:movingblk}~(right)). 
Under FBS, the track is divided into pre-defined segments or blocks, and the trains rely on track circuits to know the block occupied by the leading train. An entire block is presumed to be occupied if a train is present anywhere within the block. Moreover, a train is not permitted to enter a front block unless it is separated from the next occupied block by a threshold distance (usually fixed by the operator).
We index the blocks by $i = 1,2,\dots,$ and let $B_l(t)$ and $B_f(t)$ denote the indices of the blocks occupied by the leading and the following trains respectively at time $t$. We denote the distance of the start point of block $i$ from the origin by $d^s_i$ (see Fig.~\ref{fig:movingblk}~(right)). For safety, the trains must be separated by a minimum number of blocks, which we denote by $B_{\text{th}}.$

\begin{small}
\begin{algorithm}[!t]
	\caption{Train Motion}
	\label{alg:train_motion}
	Set $t = 0$; \\
	Compute $\mathcal{A}_{\text{plan}}(t)$ by solving \eqref{eqn:plan_profile}
	with $v_{\text{init}} = 0$ and $s_{\text{remain}} = s_{\text{tot}}.$ \\
	\While{$s_f(t) < s_{\text{tot}}$}
	{
	\uIf{$pkt\_rec = 1$} 	
	{
	Set $pkt\_loss\_counter = 0.$ \\
	$\widehat{s}_{l}(t) = s_l(t),\widehat{v}_{l}(t) = v_l(t).$ }
	\Else
	{
	$pkt\_loss\_counter = pkt\_loss\_counter+1;$ \\
	\If{$Fixed\_Blk = 0 \ \& \ pkt\_loss\_counter = N$}
	{
		$Fixed\_Blk = 1;$ \\
		$T_{\text{FB}} = T^{\max}_{\text{FB}};$
	}	

	}

	\uIf{$Fixed\_Blk = 0$}	
	{
	 Compute $H(t)$ as in Algorithm~1. \\ 
		\uIf{\textnormal{$\widehat{s}_l (t)-s_f(t) > H(t)$}}
		{
			 Set $a_f(t) = \mathcal{A}_{\text{plan}} (t).$ \\			
			 Compute $v(t+\Delta t)$ and $s(t+\Delta t)$ as in \eqref{eqn:vel_update} and \eqref{eqn:pos_update}. 
		}
		\Else
		{     Set $Fixed\_Blk = 1;$ \\
			 Perform fixed block update according to Algorithm~\ref{alg:FBS_motion}. \\ 	
			Set $T_{\text{FB}} = T^{\max}_{\text{FB}}-1;$ 		
	}
	}
	\Else{
	Perform fixed block update as in Algorithm~\ref{alg:FBS_motion}. \\
	$T_{\text{FB}} \leftarrow T_{\text{FB}} - 1;$ \\
	\If{$T_{\text{FB}} = 0$}
	{
	Set $Fixed\_Blk = 0;$ \\
	Update $\mathcal{A}_{\text{plan}}(t)$ by solving \eqref{eqn:plan_profile}
	with $v_{\text{init}} = v_f(t)$ and $s_{\text{remain}} = s_{\text{tot}} - s_f(t).$ 	
	}

		}
	Set $t \leftarrow t+\Delta{t}.$	
	}			
\end{algorithm}
\begin{algorithm}[!t]
	\caption{Fixed Block Update}
	\label{alg:FBS_motion}
	 \uIf{ $B_{l}(t)-B_{f}(t) \leq B_{\text{th}}$}
			 {
			  	Apply emergency brakes. \\
			  	Set $a_f(t) = \beta_{\text{emerg}}.$
			 }
	
			  \Else
			  {
				  Update $\mathcal{A}_{\text{plan}} (t)$ by solving  \eqref{eqn:plan_profile} with $v_{\text{init}} = v_f(t)$ and  $s_{\text{remain}} =  d^s_{B_l(t)-B_{\text{th}}} - s_f(t).$ \\
				  Set $a_f(t) = a_{\text{plan}}(t).$ \\
				Compute $v(t+\Delta t)$ and $s(t+\Delta t)$ as in \eqref{eqn:vel_update} and \eqref{eqn:pos_update}. 	
			  }
			  Return $v(t+\Delta t)$ and $s(t+\Delta t).$
\end{algorithm}
\end{small}

\subsection{Train Motion During the Operational Phase}
The train motion is described in Algorithm~\ref{alg:train_motion}.  
Before describing the algorithm, we introduce some notations used in the algorithm. Denote the time of operation by $t.$ In each time slot, we let $pkt\_rec$ denote a binary variable that indicates the status of communication during that time slot, i.e.,  it takes a value $1$ if the communication is successful, and $0$ otherwise. The variable $pkt\_loss\_counter$ counts the number of consecutive communication failures. 
Further, we use $Fixed\_Blk$ to denote a binary variable that indicates the mode of operation
of the following train during the current time slot, i.e., $1$ indicates that it is operating in FBS mode, and
$0$ indicates MBS mode. $T^{\max}_{FB}$ denotes the maximum duration that a train remains in FBS mode.

We start with the description of lines 4--12 of Algorithm~\ref{alg:train_motion}. 
During each time slot, if the communication between trains is successful, then the following train updates $\langle \widehat{s}_l(t),\widehat{v}_l(t),\widehat{a}_l(t) \rangle$ according to the latest information. Else, if there is a communication error, then $\langle \widehat{s}_l(t),\widehat{v}_l(t),\widehat{a}_l(t) \rangle$ are assumed to be the same as that of the last received information. If the number is consecutive packet losses greater $N$ (where $N$ is a pre-determined by the system operator), then it immediately switches to FBS mode. 

When in MBS mode, the following train computes a dynamic headway according to Algorithm~\ref{alg:headway} and checks to see if the distance between the trains is greater than the computed dynamic headway (line 15). If true, it continues to move according to the pre-planned guidance trajectory $\mathcal{A}_{\text{plan}}(t)$ (line 17). Otherwise, if the trains are closer than the computed dynamic headway, then for safety, it immediately switches to the FBS mode (lines 20-24). We assume that every time the train switches from MBS to FBS, it remains in FBS mode for a duration of $T^{\max}_{FB}$ in order to ensure that the separation between the trains is sufficiently large, before reverting back to MBS mode. 
If the train is in FBS mode, then it adjusts its guidance trajectory to stop
within a distance of $d^s_{B_l(t)-B_{\text{th}}} - s_f(t)$ (Algorithm~\ref{alg:FBS_motion}).
We note that $B_l(t)-B_{\text{th}}$ is the index of the block that the following train is not 
allowed to enter under FBS mode and $d^s_{B_l(t)-B_{\text{th}}} - s_f(t)$ is the distance
to the start of the corresponding block. 
After $T^{\max}_{\text{FB}}$ time slots, the train switches back to MBS mode if the packet-loss
counter is less than $N.$ Further, it recomputes the guidance trajectory for the remaining distance of its journey. The train can complete the remaining distance of its journey with minimum
time, and hence, our analysis can be viewed as a lower bound on the potential attack impact that a jammer can cause.

In the next section, we present the details of wireless communication 
model used in train-to-train or train-to-trackside infrastructure and the jamming. 

\section{Wireless Communication Channel Model}
\label{sec:Wireless}
While the greater focus is on leaky-medium-based communication channel, we also discuss the signal propagation under the free-wave channel (which corresponds to the traditional wireless medium of open air with no physical communication structure) in order to compare and contrast signal propagation and jamming in the two channels. We present the detailed description next.

{\bf Free-Medium-Based Communication:} For free-wave communication, we adopt the \emph{log-distance pathloss model} in which the path loss, $\eta$ measured in dB scale, at a distance $d$ from the transmitter is given by
\begin{align}
\eta = \eta_0 + 10 \gamma \log_{10} (d) + X, \label{eqn:PL_fw}
\end{align}
where $\eta_0$ is the reference pathloss, $\gamma$ is the pathloss exponent and $X$ is a random
variable that captures the fading effect.

{\bf Leaky-Medium-Based Communication:} The leaky-medium based communication is illustrated in Fig.~\ref{fig:Leaky_Med}.
We denote the inter-repeater distance by $d_{rptr}$ and the amplifying gain of the repeater by $C_{rptr}.$
In leaky-medium-based communication, total pathloss consists of a longitudinal component $\eta_{l},$ a radial components $\eta_{r},$ as well as the pathloss due to the repeater $\eta_{rptr},$ given by \cite{Leaky2009}, 
\begin{align}
\eta = \eta_{l}+\eta_{rptr}+\eta_{r}. \label{eqn:PL_wg}
\end{align}
The longitudinal component is linear in dB, and is given by $\eta_{l}  = C_{cplng} + \alpha d_l,$ where $C_{cplng}$ is the coupling loss and $\alpha$ is
the rate of loss over longitudinal distance $d_l.$
The radial component $\eta_{r} = \eta_{0,r} + 10 \log_{10} (d_r) + X_r$ where $\eta_{0,r}$ is
the path loss due to leakage through the slot and $X_r$ is the fading of the free wave after the leakage. Finally, $\eta_{rptr}$ is given by 
\mbox{$\eta_{rptr} = -C_{rptr} N_{rptr} ,$}
where $N_{rptr}$ is the
number of repeaters that the signal passes through. The negative sign indicates that the signal is amplified due
to the repeater (and hence the pathloss is negative).

{\bf Jamming Attack:} We consider an adversary transmitting a jamming signal with power ${P}^\prime_J$. We let ${P}^{\prime}_S$ denote the transmit power of the legitimate signal. The received powers of the legitimate and the jamming signal are denoted by $\widetilde{P}^{\prime}_S$ and $\widetilde{P}^\prime_J$ respectively.
The signal to interference noise ratio at the receiver is then given by
\begin{align}
\text{SINR} = \frac{\widetilde{P}^{\prime}_S}{\widetilde{P}^{\prime}_J + \widetilde{P}^{\prime}_N} \approx \frac{\widetilde{P}^{\prime}_S}{\widetilde{P}^{\prime}_J }, \label{eqn:SINR}
\end{align}
where in the approximation, we ignore the noise power, since we consider an interference-limited system. Jamming is successful if the SINR is below
a threshold value $\tau^\prime,$ i.e., $\text{SINR} < \tau^\prime.$ In dB scale, using \eqref{eqn:SINR} , it follows that jamming is successful if $\widetilde{P}_S - \widetilde{P}_J < \tau,$
where $\widetilde{P}_S, \widetilde{P}_J$ and $\tau$ denote the corresponding quantities in dB scale.
We note that the transmit and the received powers in dB scale are related as $\widetilde{P}_S = {P}_S - \eta_S \ \text{and} \ \widetilde{P}_J = {P}_J - \eta_J.$
Let $d_{S,R}$ denote the distance between the legitimate transmitter and the receiver and $d_{J,R}$ between the jammer and the receiver. 
In the following, we express the pathloss for the legitimate and the jamming signals under the free-medium and the leaky-medium-based communications. 

We first consider the free-medium communication. From \eqref{eqn:PL_fw}, it follows that, the pathloss for the legitimate and jammer's signals are given by
$\eta_{S} = \eta_0 + 10 \gamma \log_{10} (d_{S,R}) + X$ and $\eta_{J} = \eta_0 + 10 \gamma \log_{10} (d_{I,R}) + X$ respectively.

Next, we consider the leaky-medium-based communication. 
We follow \eqref{eqn:PL_wg} and analyze each component individually.
First, we consider the longitudinal component $\eta_l.$ 
Note that the legitimate signal from the trackside infrastructure gets injected to the waveguide by wired connection. Thus it suffers from coupling loss $C_{cplng}$ only. In contrast, the jammer's signal gets injected into the waveguide by following a wireless path. Thus, it suffers an additional pathloss due to the freewave path, given by $\eta_{J,wg} = \eta_0+10\gamma log_{10} (d_{J,wg}) + X$ (similar to \eqref{eqn:PL_fw}), where $d_{J,wg}$ is the distance of the jammer from the waveguide signal
injection point. The distance-based pathloss of the longitudinal component for the two signals
are given by $\alpha d_{S,R}$ and $\alpha d_{J,R}$ respectively. The radial component of the path loss from the leaky medium to the train receiver is constant for both the signals, which we denote by $\bar{\eta}_{r}$, as the train travels in parallel to the leaky medium and is in constant distance away in the radial direction from the leaky medium. Further, the pathloss due to the repeater for the two signals is given by $C_{rptr} N_{S,rptr}$ and $C_{rptr} N_{J,rptr}$ respectively, $N_{S,rptr}$ and $N_{J,rptr}$ denote the number of repeaters that the corresponding signals traverse through.
Based on the discussion above, we have, 
\begin{align}
\eta_{S} &= C_{cplng} +\alpha d_{S,R}-C_{rptr} N_{S,rptr}+ \bar{\eta}_{r}, \\
\eta_{J} &=  \eta_{J,wg} + \alpha d_{J,R} -C_{rptr} N_{J,rptr}+ \bar{\eta}_{r}.
\end{align}

\section{Attack Mitigation Using FHSS Repeater}
\label{sec:Mitigation}
To mitigate the jamming attack, we build on FHSS, which randomizes the frequency channel access against the attacker for jamming resistance. However, in contrast to the prior implementation of FHSS, we apply FHSS not only on the transmitter-receiver pair (the train and the track-side access point in our case) but also on the repeaters (so that the repeaters only amplify the signal going through the securely-agreed channels); the novelty comes from extending the FHSS on the wireless repeaters. FHSS implementation requires radio signal processing capabilities (requiring the radio hardware and the signal-level control, which is of finer granularity and closer to the frontend than the bit-level processor/control).

\subsection{Defense/Spreading Gain}
FHSS provides spreading gain to the legitimate train signal, which we denote by $n$. By focusing the signal power on the narrower channel, the effective SINR is increased by $n$ (since given the same power budget, the power spectral density must increase by $n$). Note that the spreading gain $n$ corresponds to the number of channels available for FHSS randomization.

While the repeater gains have been $C_{rptr}$ for all signals (legitimate and jamming) previously without FHSS, applying FHSS on the repeaters changes the repeater gain to $\frac{C_{rptr}}{n}$ for the attacker's jamming signals, assuming wideband jamming (jamming across all possible frequency channels is the optimal strategy for the attacker~\cite{simplemac_2012}). This is because the repeater filters out the rest of the channels which are not being used by FHSS at the time. In contrast, the legitimate train signal retains the repeater gain of $C_{rptr}$ because the legitimate train sends the transmission at the FHSS band.
After multiple repeaters of $N_{rptr}$, the legitimate signal gain is $(C_{rptr})^{N_{rptr}}$ and the attacker's jamming gain is $\frac{1}{n}(C_{rptr})^{N_{rptr}}$.

\subsection{Synchronization and Trust}
Per transmission symbol, the train, the access point, and the repeaters agree on a channel, and they only process the signals from that channel, effectively filtering the signals from the other channels.
Across the transmission symbols, the channels can vary.
Due to the dynamic channel operations, the legitimate entities need to agree on the channels and be synchronized in operations when the channel gets switched. Our defense builds on the prior work in FHSS (popularly used, e.g., in IEEE 802.11 legacy system) and the systems implementations development which enables the transmitter-receiver FHSS communications. In this section, we review those bases for our work.

As is typical in the traditional FHSS, the parties generate dynamic frequency hopping pattern using a pseudo-random generator (PRG), making the hopping pattern deterministic to the legitimate transmitters who have the key/seed but random to the attackers lacking the key.
Our scheme therefore assumes the key distribution for the seed driving the PRG, e.g., using prior work~\cite{hartong2006,kuts2016}.

Our scheme also requires frequency-channel synchronization in time across the participating nodes of the train, the access point, and the repeaters.
While wireless systems use time-interleaved beacon/preamble signal for time synchronization (for example, by appending the preamble signal before the data signal), we propose the use of synchronization signaling that is independent of the channel selected for FHSS (e.g., out-of-band signaling) to thwart reactive jamming (which senses the spectrum use in real time and adapt the jamming strategy accordingly). 
The synchronization problem is generally easier than in the context of the multiple wireless nodes in a distributed setting because of the more tightly controlled train networking environment, e.g.,
the OCC (which is consistently communicating with the train) controls the repeaters as part of its infrastructure
and the train systems can afford to train and calibrate the networking and the corresponding operations regularly, e.g., when the trains are not in operations.

The train-to-infrastructure communications correspond to many-to-one communication in the uplink, as there can be multiple trains in operations while there are smaller number of access points (all of which are connected to one centralized OCC). Therefore, there can be signal collisions from multiple trains, and one train's transmission can interfere with another. To support multiple coexisting transmissions, wireless communications provide multiple channels in time, frequency, code (e.g., direct-sequence spread spectrum (DSSS) and code-division multiple access (CDMA)), and in space (e.g., multiple-input multiple-output (MIMO)-based beamforming and picocell/microcell). To provide greater efficiency in such channel use as a network, wireless systems also make use of medium access control (MAC) protocols which have the subset of the network users agree on the channel use and broadcast the channel-agreement information to other users to avoid collisions. While both accidental collisions and jamming can result in destructive interference to the communications receiver, jamming poses a worse-case interference source and a greater problem because of the malicious nature of the jamming source (we assume that the jammer's goal is to disrupt the communications); therefore, while such wireless MAC-layer measures may be effective for accidental interference, we build on them for general interference resistance and use FHSS to defend against jamming.

\section{Simulations: Jamming Attack Impact}
\label{sec:Sim_Res}

\subsection{Simulation Settings and Methodology}
The simulations are carried out in MATLAB. All the constrained optimization problems in the simulations are solved using the \emph{fmincon} function of MATLAB.

We simulate the motion of multiple trains along a single metro line consisting of $30$ stations. Each train commences its journey from Station~1 and ends at Station~30. The trains are continuously dispatched with a fixed dispatch interval of $90~$seconds starting
from 8:00:00~AM. Between stations, trains run according to the motion profile described in Section~\ref{sec:Train_Motion}, and stop for a duration of $30~$seconds at each station.
For the train passenger flow, we use the dataset provided by the Shenzhen metro line \cite{Zhang:2015}.
We use the data corresponding  to the ``Green line" (which has $30~$ stations), starting from 8:00:00~AM until 10:00:00~AM. We serve the passengers in this data-set by trains running according to our simulations. (The dataset provides passenger smart card tap-in and tap-out times, which enables us the determine their origin and destination stations, as well as their arrival times at different stations.) 

The train motion parameters are chosen as follows. 
The acceleration/declaration values are set to $\alpha = 0.7 \ \text{m}/s^2,$ $\beta_{\text{ser}} = 0.4 \ \text{m}/s^2,$ and $\beta_{\text{emerg}} = 1 \ \text{m}/s^2.$ The maximum train velocity is $v_{\max} = 16.67 \ \text{m}/s$ (i.e., $60~$km/hr). The inter-station distance between two adjacent stations is taken to be $2.8~$kms for all the stations. For fixed block, the block length is set to $400~$m. We note that these simulation parameters approximately reflect the settings of a real metro system. 

The interval between train operational decisions (i.e., $\Delta t$ defined in Section~\ref{sec:Train_Motion}) is assumed to be $0.25~$seconds. 
We assume that if communication failure occurs for a duration $2~$seconds, then the train switches from MBS mode to FBS mode. Thus $N = \frac{2}{\Delta t} = 8$ (recall its definition from Section~4.3). Further $T^{\max}_{\text{FB}}$ is set to $30~$seconds.

Next, we list the wireless communication parameters.
The SINR threshold for successful communication is set to $\tau = 10~$ dB.
For the free-medium communication, the reference pathloss $\eta_0$ is taken to be
$90~$dB. For simplicity, we ignore the fast fading component $X_r$ in our simulations.
For the leaky waveguide simulation parameters, we adhere to the Electronic
Industries Alliance Waveguide WR-430 standards \cite{Leaky2009}. Accordingly, the parameters
are chosen as $C_{cplng} = 0.3~$dB, $\alpha = 17~\text{dB}/km,\bar{\eta}_r = 62~$dB, $C_{rptr} = 42.5~$dB, $\gamma = 2.$ The inter-repeater distance $d_{rptr}$
is set to $2.5~$kms. (This distance is selected such that in the absence
of the jammer, the signal-to-noise ration SNR of legitimate communication signal
is always greater than the threshold $\tau$.)
The transmit powers of the legitimate signal and jammer, $P_S$ and $P_J,$ are taken to be $23~$dBm each.
The attacker is assumed to be located at a distance of $0.2~$km from
the origin and continuously transmits the jamming signal throughout the simulation interval.

The passengers are served as follows. Whenever a train reaches a particular station, all the passengers who have arrived at the station before that time (in the dataset) are allowed to board the train, subject to a train capacity constraint, which is assumed to be $400$ passengers. If train
capacity is reached, the remaining passengers wait for subsequent trains. The total passenger journey time is the sum of the wait time, (i.e., the time when he/she arrives at the station according to the tap-in time listed in the data-set and the time he actually boards the train), and the train journey time till his/her destination station. 
The simulation starts when the first train commences its journey from Station~$1$, and is carried out for a duration of $2~$hours (i.e., 8:00--10:00 AM). Whenever a train reaches the destination Station~$30$, it is removed from the simulation.

\subsection{Simulation Results}
\emph{Comparison of Jamming In Leaky-Medium and Free-Medium Communication:} 
\begin{figure}[!t]
	\centering
	\includegraphics[width=0.48\textwidth,trim={0 8.7cm 0 0}]{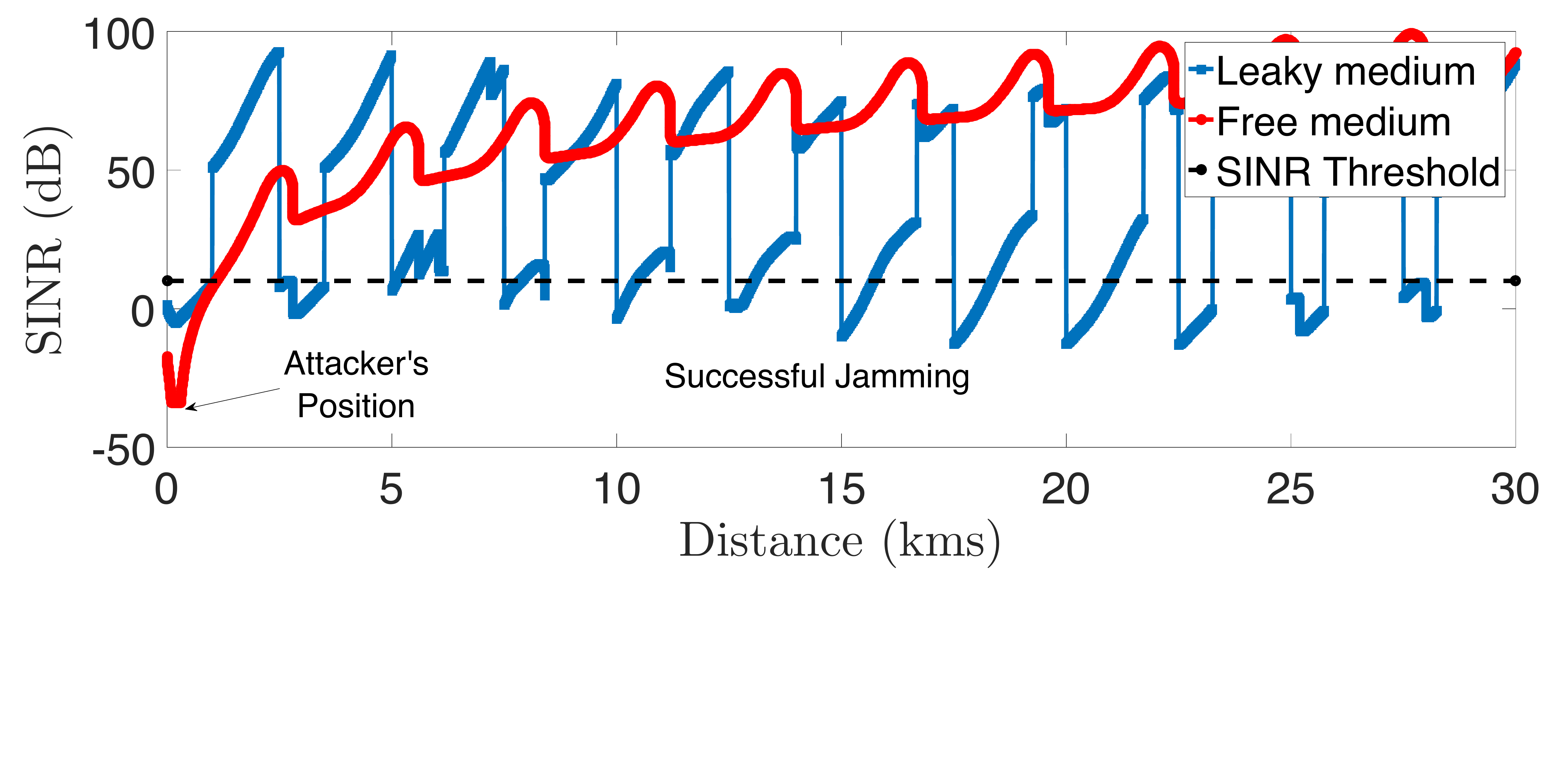}
	\caption{SIR For Leaky Wave And Free Wave Jamming.}
	\label{fig:SIR_FW_Leaky}
\end{figure}
We first compare and contrast jamming in leaky-medium communication against jamming in free-medium communication. 
In Fig.~\ref{fig:SIR_FW_Leaky}, we plot the SINR for one of the trains ($14^{\text{th}}$ dispatched train) as a function of its position with respect to the origin in the two settings. The SINR threshold and attacker's position are also marked in the figure. It can be observed that for the free-medium communication, jamming is effective only over a limited range, i.e., when the train is in proximity of the jammer. In contrast, for the leaky-medium communication, jamming is effective throughout the train communication space. This can be explained as follows. Recall that the pathloss suffered by the legitimate signal depends on the distance between the trains, where as the pathloss suffered by the
jammer's signal depends on the distance between the jammer and the receiver (i.e., following train). In case of free-medium communication, the intensity of the jamming signal degrades as the train moves away from the jammer. 

\begin{figure}[!t]
\centering
\includegraphics[width=0.48\textwidth,trim={0 1cm 0 0}]{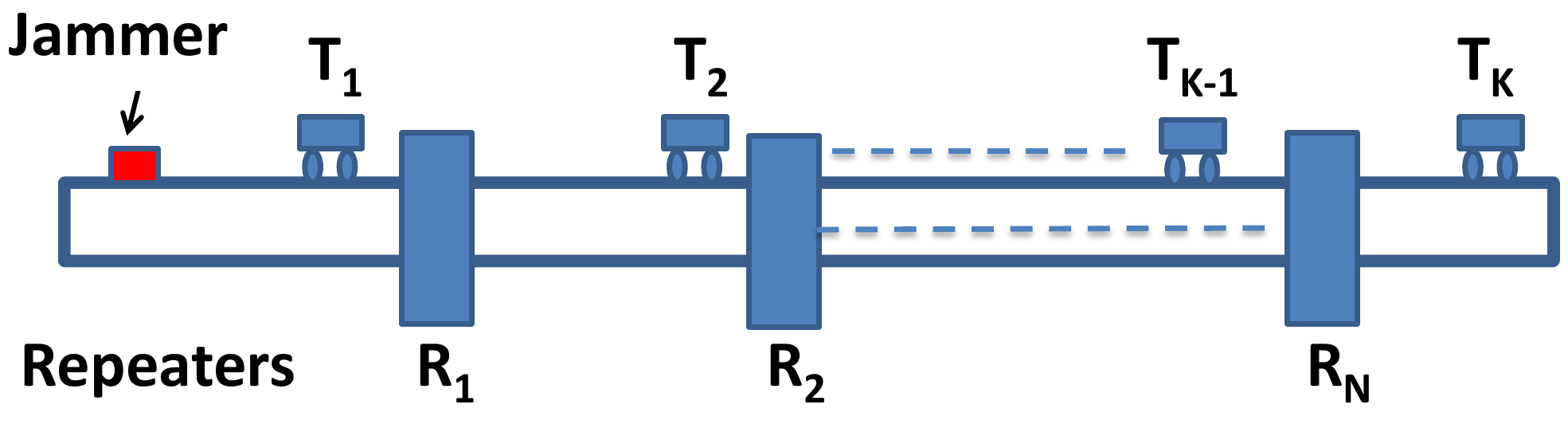}
\caption{Simulation setup.}
\label{fig:Setup}
\end{figure}

In contrast, for the leaky medium, we observe that the SINR exhibits a periodic pattern and the attacker is able to successfully jam the train communication multiple times. This can be explained as follows. In Fig.~\ref{fig:Setup}, consider two scenarios: Scenario~1 in which train T$_1$ is just behind the repeater R$_1$ and Scenario~2 in which T$_1$ has just passed R$_1$. It can be noted that in Scenario~2, the jammer's signal receives an additional amplification of $C_{{rptr}}$ units compared to Scenario~1 (as its signal passes though the R$_1$). Thus, the SINR in Scenario~2 is significantly lower than Scenario~1, and hence more favourable for the jammer. As the train T$_1$ moves away from the repeater, the jammer's signal intensity starts decreasing due to the path loss, due to which the SINR increases. This explains the periodic pattern in the train's SINR.

\begin{figure}[!t]
\centering
\includegraphics[width=0.48\textwidth,trim={0 9.8cm 0 0}]{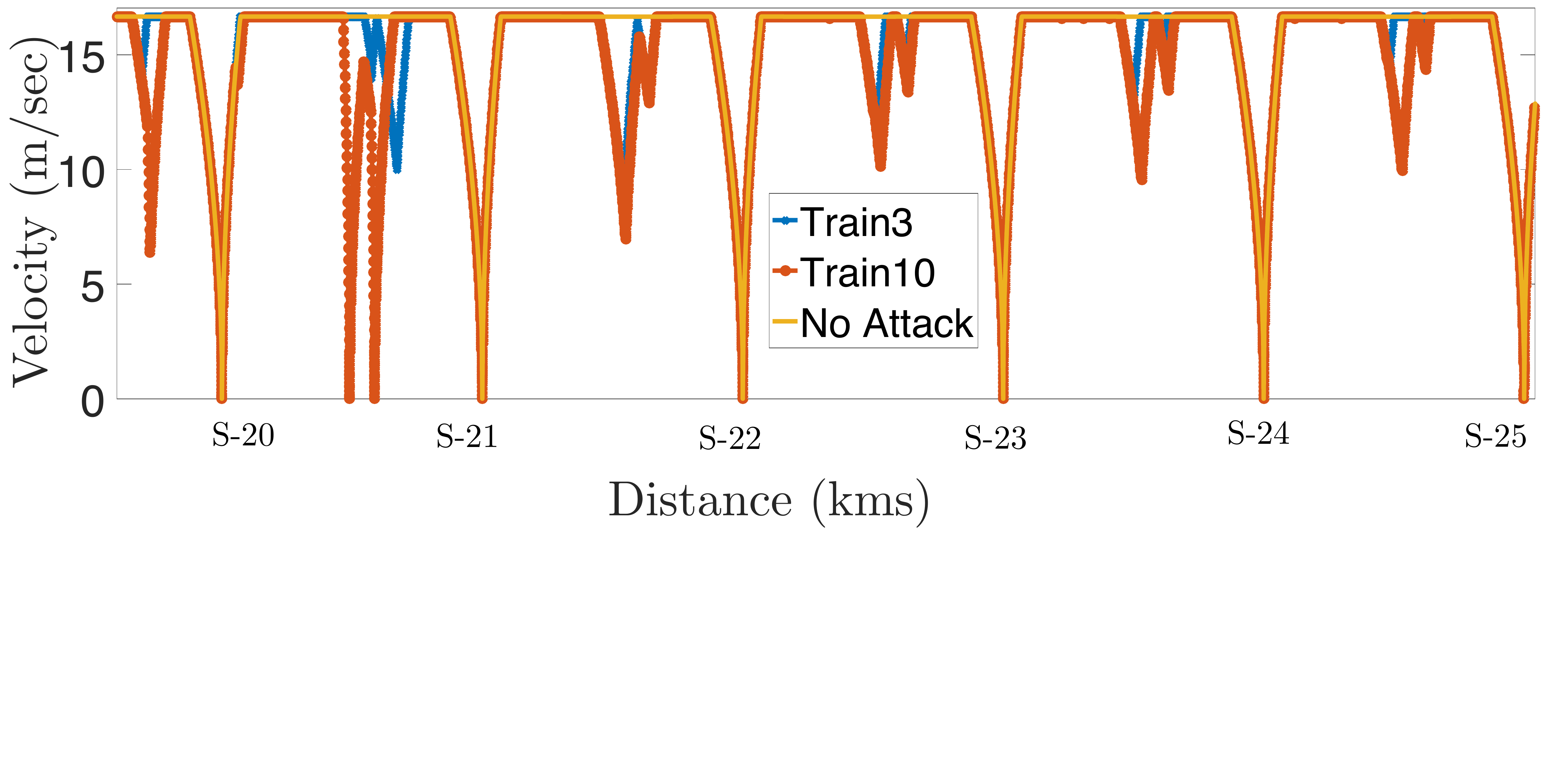}
\caption{Train velocity profile with leaky-medium communication.}
\label{fig:Leakywave_Jam}
\end{figure}
\emph{Jamming Impact on Train Motion \& Passenger Flow:}
Next, we investigate the jamming impact using the co-simulation approach. Unless mentioned otherwise, we present simulation results considering the leaky-medium communication only as jamming is more feasible and impactful 
in this medium than the open-air free medium (refer Fig.~\ref{fig:SIR_FW_Leaky}). We plot the velocity of two trains (i.e., train~$3$ and train~$10$) as a function of their position in Fig.~ \ref{fig:Leakywave_Jam} between stations $20$ and $25$. The yellow curves indicate the train velocities during normal operation (without the jammer), where as, the red and blue curves indicates the train velocity with jamming. It can be observed that in the presence of the jamming attack, the train brakes frequently (observe the reduction in train velocity). This is due to loss of signal, due to which they frequently switch to FBS mode. Consequently, the train has to decelerate in order to conform to the fixed block headway.

We tabulate the train journey time with and without jamming in Table~\ref{tbl:FW_delay} in Appendix~C.
It can be noted that the increase in train journey time for the leaky-medium communication is significantly higher than the free-medium communication. 
Moreover, the overall increase in train journey time for free-medium communication is negligible due to the limited jamming.  We also observe a cascading effect in the jamming attack impact for leaky-medium communication, i.e., the attack impact is higher on the trains that are dispatched later. For instance, the train dispatched at 08:06:00~AM suffers an increase $13.6 \%$ increase, where as the train dispatched at 09:52:30~AM suffers an increase of $34.15 \%.$ This is because the trains slow down due to jamming, and its effect propagates over successive trains and becomes more severe.

\begin{figure}[!t]
\centering
\includegraphics[width=0.48\textwidth,trim={0 8cm 0 0}]{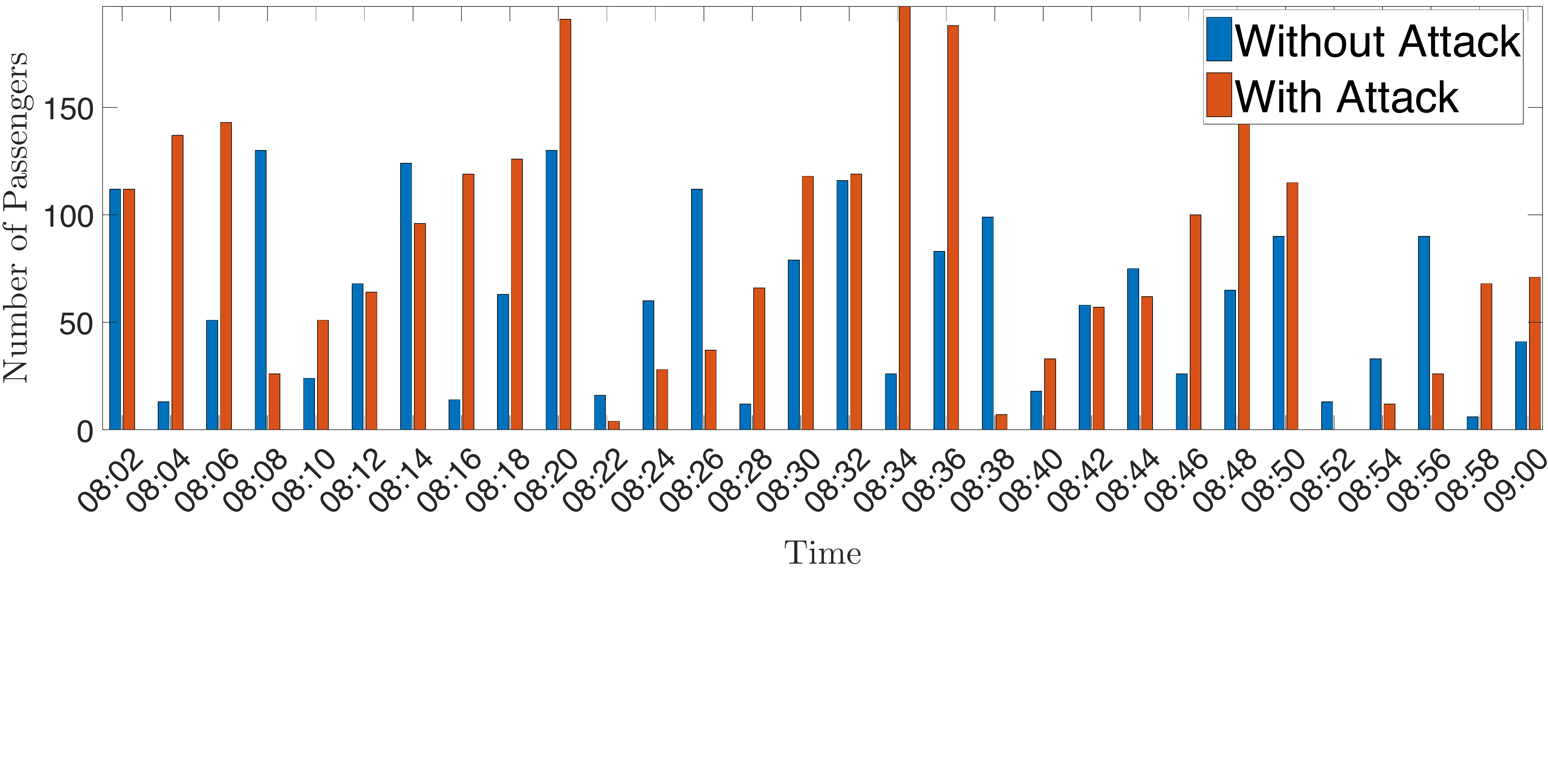}
\caption{Passenger congestion at Station~3 between 9--10~AM with and without attack.}
\label{fig:Station_Congestion}
\end{figure}

We also investigate the attack impact in terms of the passenger travel time and station congestion.
To the illustrate station congestion, we plot the number of passengers waiting at Station~3 between $9-10$~AM in Fig.~\ref{fig:Station_Congestion}. It can be observed that the passenger congestion significantly increases with the attack, which can lead to an increase in the customer wait time.

Overall, between $9-10~$AM, we observed that on an average, the passengers suffer from  $33 \%$ increase in their total journey time. We also observed that the attack impact was more severe on passengers whose original journey time (i.e., without attack) was long. For instance, passengers whose journey time without attack was greater than $40~$mins suffered from an increase of about $15~$mins or more due to the attack. This is due because trains move slower due to jamming, which resulted in an increase in the overall journey time as well as the station congestion (which in turn can potentially increase the passenger wait time).

\section{FHSS Repeater Prototype and Its Efficacy}

\label{sec:FHSS_proto}
We develop a prototype for FHSS repeater, our defense for jamming mitigation as described in Section~\ref{sec:Mitigation} to demonstrate the effectiveness of our scheme.
In this section, we first present the details of the SDR-based prototype and the testbed environment using leaky feeder/coaxial cable system and then show the experimental results using the prototype testbed. In the second part, we 
integrate the FHSS mitigation technique in our co-simulation to investigate its effectiveness 
 in limiting the jamming attack impact.


\begin{figure}[!t]
\centering
\begin{subfigure}{0.23\textwidth}
\includegraphics[width=1\textwidth]{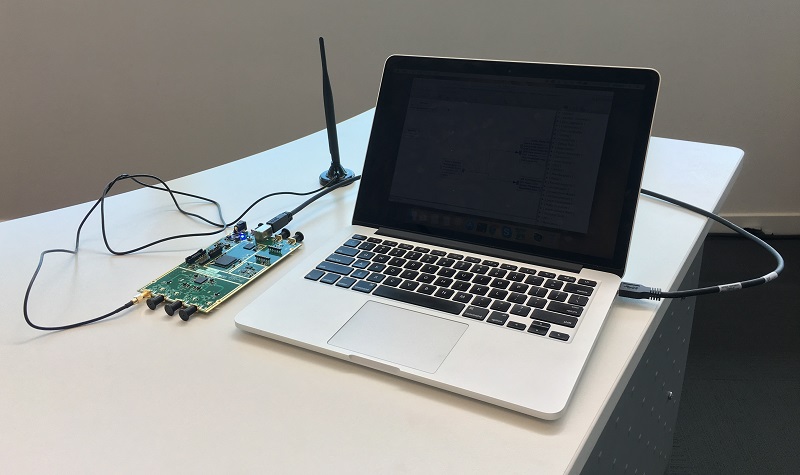}
\end{subfigure}
~
\begin{subfigure}{0.23\textwidth}
\includegraphics[width=0.9\textwidth]{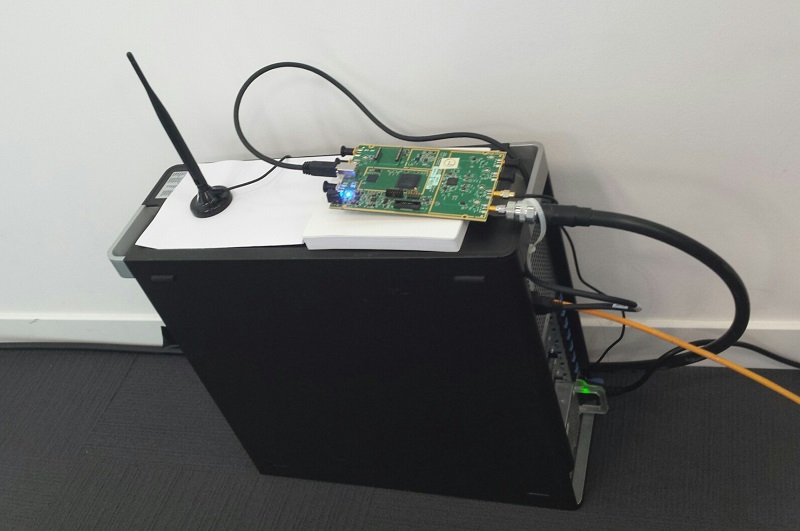}
\end{subfigure}
\caption{Device setup for the experiments (left) and device setup for the repeater node (right).}
\label{fig:Exp}
\end{figure}

\begin{figure}[!t]
\centering
\begin{subfigure}{0.23\textwidth}
\includegraphics[width=0.98\textwidth]{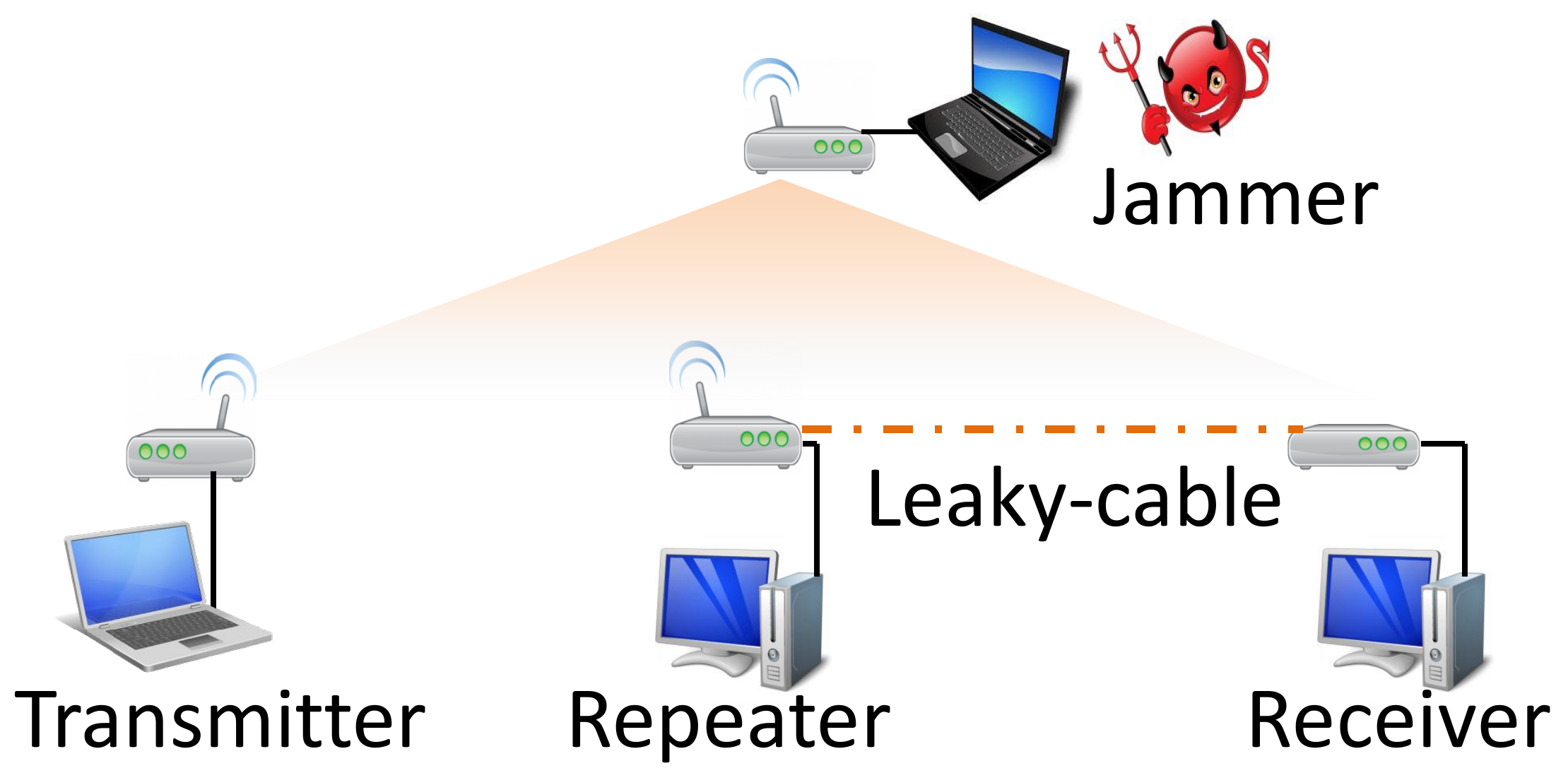}
\end{subfigure}
\begin{subfigure}{0.23\textwidth}
\includegraphics[width=0.98\textwidth]{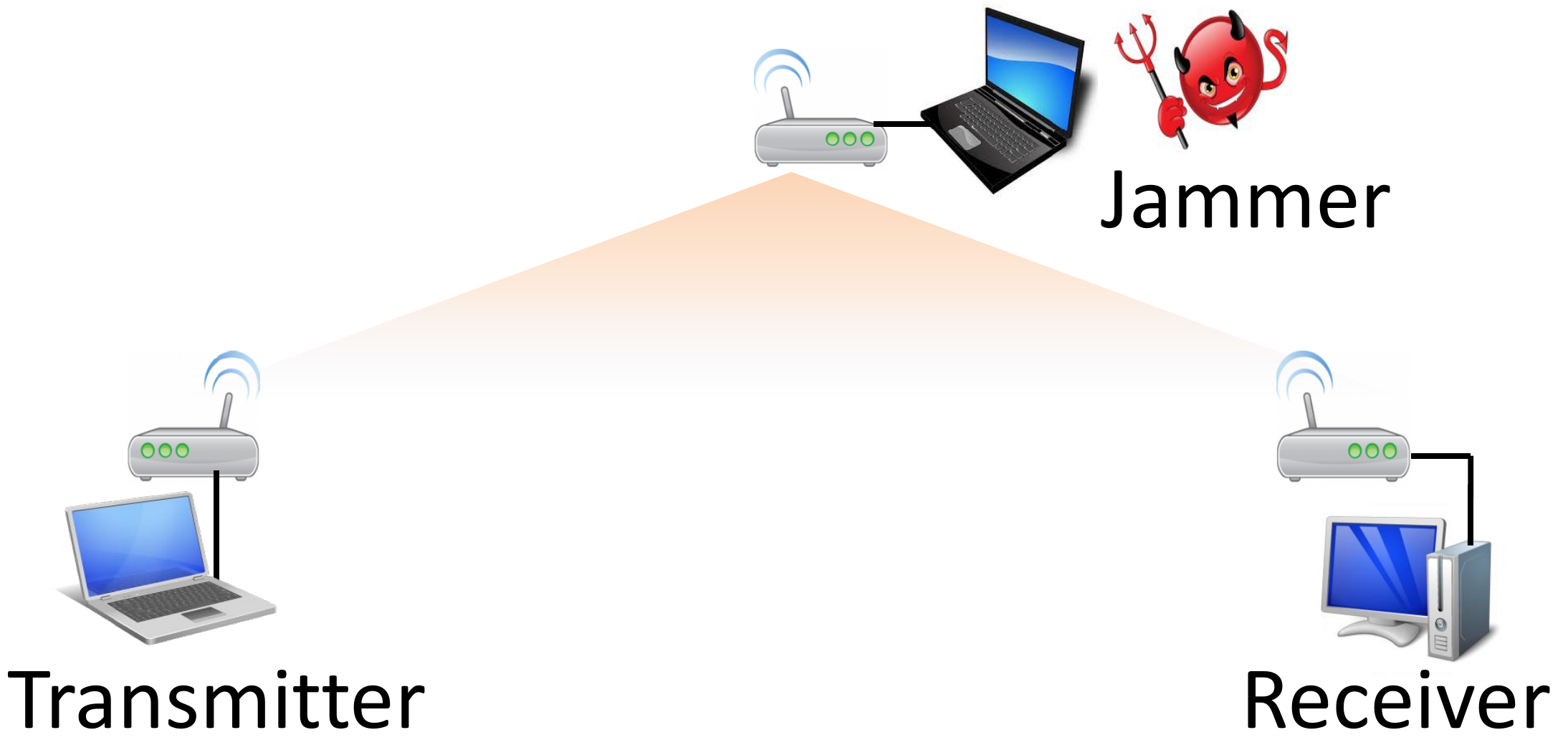}
\end{subfigure}
\caption{Experimental setup for leaky-medium (left) and free-medium (right) communication.}
\label{fig:expt_cable}
\end{figure}

{\bf FHSS Repeater Prototype:} We first present details of the FHSS repeater prototype. 
Each node is prototyped using a Universal Software Radio Peripheral (USRP) B2100 board~\cite{usrp} hosted and processed by a computer, as shown in Fig.~\ref{fig:Exp}~(left). We implement the functionality of each nodes using GNURadio \cite{gnuradio}. 
The testbed comprises of four nodes: the transmitter (Tx), the receiver (Rx), the repeater (Rr), and the jammer (J).

\textit{i) Setup for leaky-medium communication:} 
A schematic diagram for this scenario is shown in Fig.~\ref{fig:expt_cable}~(left). 
The Tx node, that emulates a train, consists of an omni-directional antenna transmitting into the free-wave medium. 
The Rr node receives the Tx signal 
using an omni-directional antenna and the Rr retransmits the signal
into the leaky coaxial cable using a direct connection (the repeater setup is shown in Fig. ~\ref{fig:Exp}~(right)).
The Rx node is also directly connected to the leaky coaxial cable and receives the signal sent from the Rr.
The jammer node consists of an omni-directional antenna, which injects its signal into the free-wave medium. This signal in turn 
gets injected into the leaky coaxial cable.

\textit{ii) Setup for free-medium communication:} A schematic diagram for this scenario is shown in Fig.~\ref{fig:expt_cable}~(right). 
In this setup, both the Tx and Rx employ omni-directional antenna and communicate over the free wave. There is no Rr node in this scenario. The antenna gains of Tx, and Rx are set to same values as in the leaky-medium setup for comparison. The jammer also consists of an omni-directional antenna transmitting into the free wave. 

\emph{Experiments and Results:} In our experiments, all the wireless transmissions take place between $400$ to $400.5~$MHz (we choose this particular frequency band because it was free of interference in our laboratory environment.). 
For simplicity, we transmit/receive analog signals only (hence we do not perform modulation and coding operations in our experiments).
The legitimate nodes (Tx, Rx, and Rr) employ FHSS for their communication and use a common PRG to 
decide the channel hopping pattern. The PRG seed is securely communicated to them by a central server (PC in our case, which emulates the role of a central OCC in railways). 
The server also broadcasts a control command to the legitimate devices to initiate the channel hopping. The hopping duration is $1$ second before switching it to another channel.

For each measurements, we repeat our experiment with $10$ different PRG seeds, and with each seed, we perform channel hopping $100$ times for sampling.
The Tx, Rx and Rr log their channel hopping details to a file separately for post-processing and analysis. We also verify correct and reliable communications in the absence of jammer.
When the jammer is present, it performs wideband jamming by uniformly distributing its power on all the channels, as it is the jammer-optimal strategy in SINR and channel capacity~\cite{simplemac_2012} given a finite power budget.
All nodes including both the legitimate and the jammer have the same power budget, and we manually calibrate the antenna gains and the node locations in order to match the SINR close to $0~$dB when using 1 channel in the free-wave environment, so that the jammer successfully disrupts the communications if the SINR threshold $\tau > 0$. 

\begin{figure}[!t]
	\centering
	\includegraphics[width = 0.45\textwidth,trim={0 0.5cm 0 0}]{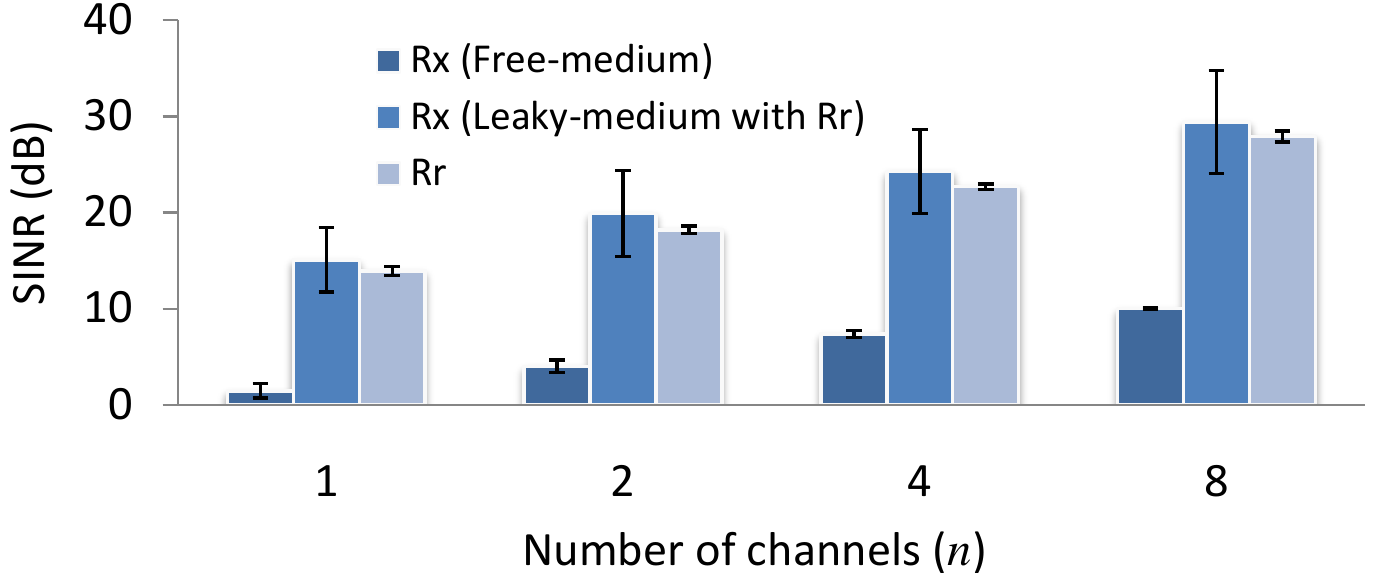}
	\caption{Effect of increasing the number of channels on SINR.}
	\label{fig:sinr_rptr}
\end{figure}

To evaluate the spreading gain and the corresponding jamming resistance, we vary the number of hopping channels available for FHSS ($n$) while fixing the bandwidth for each channels to be $50~$kHz, i.e., the total bandwidth is $50 \times n$ kHz. Fig.~\ref{fig:sinr_rptr} varies $n$ and presents three SINR measurements (the means and the confidence intervals): the SINR at the receiver using the free-wave channel (``Rx (Free-medium)''), the SINR at the receiver using the leaky-medium channel and an intermediate repeater (``Rx (leaky-medium with Rr)''), and the SINR measurement at the repeater (``Rr''); the first corresponds to the setup described in Fig.~\ref{fig:expt_cable}~(right) while the latter two is experimented in the setup in Fig.~\ref{fig:expt_cable}~(left). In all cases, the SINR increases as $n$ increases due to the FHSS spreading gain~\cite{pickholtz1982,simon1994}; the attacker spreads its power and therefore its effective interference power at the receiver or the repeater decreases as $n$ grows. The SINR grows proportionally to $n$ in free-wave channel, which corroborates with prior literature in spreading spectrum, while the spreading gain $n$ provides even greater gain in leaky-medium channel with a repeater.

Comparing the free-wave channel and the repeater-aided leaky-medium channel, the leaky channel with the repeater outperforms the free-wave medium for both the receiver and the repeater consistently across $n$ because the leaky medium offers guided propagation whereas free-wave does not and can be subjected to multi-path effects.
In the leaky-medium communication environment, the receiver SINR is greater than the SINR at the intermediate repeater because of the repeater gain $C_{rptr}$, which was fixed to be $70~$dB for these measurements. 
However, while the repeater presence increases the expected SINR because of $C_{rptr}$, 
using a repeater increases the variance/randomness, as indicated by the confidence interval; the confidence interval is larger for the receiver with the repeater and the leaky-medium channel than other SINR measurements.

The choice of $n$ offers a tradeoff between reliability/SINR performance and the bandwidth use (which is a valuable resource in wireless communications). Choosing $n$ when implementing FHSS repeater depends on the nodes' relative power (and especially that of the jammer's) and the SINR threshold $\tau$ (which is dependent on the physical-layer processing of modulation and coding, e.g., redundancy and error correcting code). For example, if $\tau=5~\text{dB}$, then reliable communication requires $n\geq 4$ for free-wave channel while it is sufficient for $n=1$ for leaky medium with one repeater; the free-wave channel consumes four times as much bandwidth for the transmission as the leaky-medium channel in this case. 
If the SINR requirement increases and is $\tau=10~\text{dB}$ (as in the case in Section~\ref{sec:Sim_Res}), e.g., the physical-layer processing at the receiver is more aggressive for greater data goodput and has less redundancy and error-correcting, then reliable communication requires $n> 8$ for free-wave channel and still $n=1$ for leaky-medium channel, in which case the bandwidth consumption for free-wave is more than eight times of that of the leaky medium.

\begin{figure}[!t]
\centering
\includegraphics[width=0.42\textwidth,trim={0 2cm 0 0}]{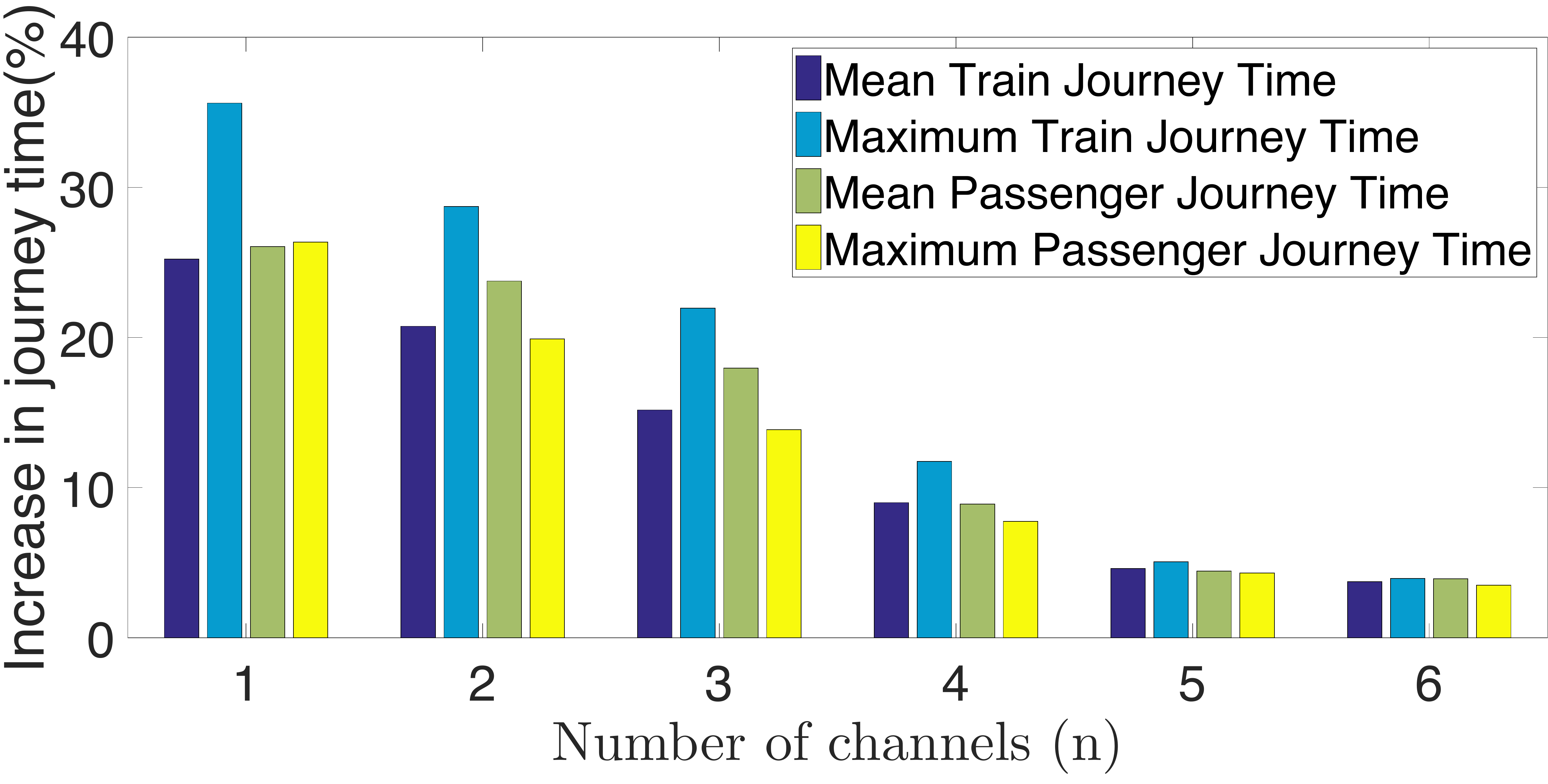}
\caption{Percentage increase in the train and the passenger journey time (with respect to their values without attack) under leaky-medium communication as a function of the number of available channels $n$.}
\label{fig:Journey_Times}
\end{figure}

{\bf Efficacy of FHSS Mitigation:} We incorporate FHSS mitigation into our simulations. 
The simulation settings are identical to that of Sec.~\ref{sec:Sim_Res}.1 (including the transmission
powers of the legitimate signal and the jammer as well as the jammer's position).
We plot the percentage increase in train journey time (with respect to their values without attack) as a function of the number of channels $n$ available for train communication in Fig.~\ref{fig:Journey_Times} for the trains whose start time is between 08:00:00~AM and 10:00:00~AM. We only consider the leaky-medium communication in our simulations as jamming is impactful only in this medium (refer Sec.~\ref{sec:Sim_Res}).  
It can be observed that the proposed FHSS strategy significantly mitigates the jamming attacks, and hence, the trains can operate under moving block mode for longer durations. Consequently the attack impact becomes almost negligible for $n = 10$ channels.

We also evaluate the impact on passenger's total journey time (sum of the wait time and the travel time). It is observed that the total passenger journey time also drops down with increase in number of channels. The waiting time of passengers reduces as the trains arrive at the stations more frequently and there are less number of denied boardings. Moreover, the travel time of passengers also reduces due to the FHSS mitigation. (We note that the total journey time of the passengers is different from the train journey time as it also includes the waiting time at the station.)

\section{Conclusions}
\label{sec:Conc}
In this work, we investigated the end-to-end impact of signal jamming attacks against CBTC systems in terms the increase in train journey time and the passenger wait time/congestion using a co-simulation approach involving model of the train motion under different signaling modes (MBS and FBS mode) and the wireless communication channel under free-medium and leaky-medium communication.
Our results show that jamming can have a particularly severe impact in the leaky-medium-communication based CBTC system by leveraging on
the signal amplifying aspect of the repeaters. 
To mitigate the attack, we proposed a FHSS strategy and evaluated the proposed solution with an SDR-based testbed. Our results demonstrated that FHSS method significantly improve the SINR at the receiver for both free-medium and leaky-medium communication methods and
effectively limit the attack impact.

\newpage
\bibliographystyle{ACM-Reference-Format}
\bibliography{bibliography}

\section*{Appendix~A: Train Motion Profile}
In this appendix, we present an optimization formulation to compute the duration of acceleration, cruising, coasting and braking phases (i.e., $T_{1},T_{2},T_{3}$ and $T_{4}$ respectively) during the planning phase of a train's motion with an objective 
of minimizing the train's total journey time\footnote{We choose this objective since the train journey time (and the corresponding passenger congestion) is the primary metric of interest in this work.}. It can be cast as follows:
\begin{subequations}
\label{eqn:train_profile}
\beqa
& \dsp  \min &  T_{1} + T_{2} + T_{3} + T_{4}   \\
&s.t.  & v_{{1}} = v_{\text{init}}  + \alpha  {T_{1}},  \label{eqn:1b}\\
& & v_{2} = v_{1} + a_{fr} T_{3}, \\
& & v_2 + \beta_{\text{ser}} T_4 = 0 \label{eqn:1p}\\ 
& & {s_{1}} = v_{\text{init}} T_1 + \frac{1}{2} \alpha {T^2_{1}} , \label{eqn:1c} \\
& & {s_{2}} = v_{{1}}{T_{2}},   \label{eqn:1d} \\
& & s_{3} = v_{1}{T_{3}} + \frac{1}{2} a_{fr} {T^2_{3}} ,\\
& & {s_{4}} = v_{2}{T_{4}} + \frac{1}{2} \beta_{\text{ser}} T^2_{4} , \label{eqn:1e} \\
& & s_{1}+s_{2}+s_{3}+s_{4} = s_{\text{remain}},  \label{eqn:1f} \\
& & 0 \leq v_{2} \leq v_{1} \leq v_{\max},   \label{eqn:1g} \\
& &  T_{1},T_{2},T_{3},T_{4} \geq 0 ,  \label{eqn:1h} \\
& \text{{\bf Inputs:}}  & \ v_{\text{init}}, s_{\text{remain}},\alpha,\beta_{\text{ser}}, a_{fr}, \nonumber \\
 & \text{{\bf Outputs:}} & \ T_{1},T_{2},T_{3},T_{4},v_{1},v_{2}. \nonumber
\eeqa
\end{subequations}
In \eqref{eqn:train_profile}, $v_{\text{init}},v_{1}$ and $v_{2}$ are the initial, cruising and coasting velocities respectively, $s_1,s_{2},s_{3}$ and $s_4$ are distances travelled during the corresponding phases. Constraints \eqref{eqn:1b}-\eqref{eqn:1e} are the velocities and distances travelled during the these phases computed according to Newton's laws of motion.
 The total distance
travelled must be equal to the remaining distance of the journey, $ s_{\text{remain}}$ (constraint \eqref{eqn:1f}). Constraint \eqref{eqn:1p} implies that the train must come to rest when
it reaches the following station.
When the train starts its journey, we have that, $v_{\text{init}} = 0,$ and
$ s_{\text{remain}} =  s_{\text{tot}},$ where $s_{\text{tot}}$ is the total distance between two consecutive stations. Note that our model ignores some finer details such 
as the constraints due to track conditions (e.g., its gradient and curvature) and train's jerk, which can be easily incorporated. 

The train's guidance acceleration profile, which we denote by $a_{\text{plan}}(\tau),$ is then given by
\begin{small}
\begin{align}
& a_{\text{plan}}(\tau) = 
&\begin{cases}
	\alpha, & t \leq \tau \leq t+T_{1}, \\
	0, &  t+T_{1} < \tau \leq t + T_{1} + T_{2}, \\
    a_{fr}, &  t+ T_{1} + T_{2} < \tau \leq  t+ T_{1} + T_{2}+T_{3}, \\
	\beta_{\text{ser}}, & t+ T_{1} + T_{2}+T_{3}  < \tau \leq t+ T_{1} + T_{2}+T_{3} + T_{4},
\end{cases} \label{eqn:plan_profile}
\end{align}
\end{small}
where $t$ is the time index when the guidance trajectory is computed ($t = 0$ at the beginning to train's journey).

\section*{Appendix~B: Dynamic Headway Computation Under Moving-Block Signalling}
In MBS, dynamic headway is the minimum separation between the two trains to avoid collision under the worst-case stopping scenario (i.e., when the leading train applies emergency brake). Note that the leading train communicates its velocity and position (i.e., $\widehat{v}_l(t),\widehat{s}_l(t)$) to the following train via the trackside equipments. Based on this information,
the following train computes the dynamic headway as follows:
\begin{small}
\begin{algorithm}
	\caption{Moving Block Headway Computation}
	\label{alg:headway}
	{\bf Inputs:} $t,\widehat{v}_l(t),\widehat{s}_l(t),v_f(t),s_f(t).$\\
	{\bf Output:} $H(t).$\\
	Set $H(t) =0, \widetilde{v}_f = v_f(t), \widetilde{v}_l = \widehat{v}_l(t),\widetilde{s}_f = s_f(t).$
	\While{$\widetilde{v}_f \geq 0$}
	{	$\widetilde{s}_{l}  \leftarrow \widetilde{v}_{l} \Delta{t} + \frac{1}{2} \beta_{\text{emerg}}(\Delta{t})^2, \widetilde{s}_{f}  \leftarrow \widetilde{v}_{f} \Delta{t} + \frac{1}{2} \beta_{\text{ser}} (\Delta{t})^2$ \\
		$ \widetilde{v}_{l} \leftarrow \max(\widetilde{v}_{l} + \beta_{\text{emerg}}\Delta{t},0), 
		\widetilde{v}_{f} \leftarrow \widetilde{v}_{f} + \beta_{\text{ser}}\Delta{t}$ \\
$H(t) \leftarrow \max( \widetilde{s}_f  - \widetilde{s}_l,H(t)).$
	}
\end{algorithm}
\end{small}

In the above algorithm, steps~4--6 ensure that the following train has sufficient distance to stop using its service brake when the leading train applies the emergency brake (the emergency brake deceleration is denoted by $\beta_{\text{emerg}},$ where $\beta_{\text{emerg}} > \beta_{\text{ser}}$).

\section*{Appendix C: Simulation Results}
In this appendix, we present the total train journey time of a few of the trains from
our simulations.

\begin{table}[!h]
\centering
\begin{tabular}{ |c|p{1.6cm}|c|p{1.6cm}|c| } 
\hline
\multirow{2}{*}{Start time} &\multicolumn{2}{|c|}{Leaky-medium} &\multicolumn{2}{|c|}{Free-medium}\\
\cline{2-5}
& Journey time (mins)& \%Inc & Journey time (mins)& \%Inc\\ 
\hline
08:06:00 & 131.22&15.71 & 113.4&0.02\\ 
\hline
08:45:00 & 138.63&22.27 &113.4&0.02\\
\hline
09:52:30 & 152.1& 34.15 &113.4&0.02\\
\hline
\end{tabular}
\caption{Train journey time under jamming with free and leaky medium. Journey time without attack 
is $113~$mins.}
\label{tbl:FW_delay}
\end{table} 

\end{document}